\documentclass{PoS}
\usepackage{graphics}
\usepackage{amsmath}
\usepackage{amssymb}
\usepackage{bbm}
\usepackage{upgreek}
\usepackage[position=t,caption=false]{subfig}

\newcommand{\absvec}[1]{\left| \vec{#1} \right|}
\DeclareMathOperator{\tr}{tr}
\DeclareMathOperator{\real}{Re}
\newcommand{\identity}{\mathbbm{1}}

\newcommand{\unit}[1]{\ensuremath{\mathrm{\,#1}}}
\newcommand{\mass}[3]{\mathrm{#1}^{#2} {#3}\;}
\newcommand{\euler}{\mathrm{e}}
\newcommand{\iunit}{\mathrm{i}}
\newcommand{\deltafunction}{\updelta}
\newcommand{\hati}{\hat{\imath}}

\title{Coulomb gauge on the lattice: From zero to finite temperature}

\ShortTitle{Coulomb gauge on the lattice: From zero to finite temperature}

\author{\speaker{Hannes Vogt}, Giuseppe Burgio, Markus Quandt, Hugo Reinhardt
         \\
          Eberhard Karls Universit\"at T\"ubingen, Institut f\"ur Theoretische Physik\\
        E-mail: \newline
        \email{hannes.vogt@uni-tuebingen.de},
 \email{giuseppe.burgio@uni-tuebingen.de},
\email{markus.quandt@uni-tuebingen.de},
\email{hugo.reinhardt@uni-tuebingen.de}\\
}
        
\abstract{Our previous studies of Coulomb gauge Yang-Mills theory are extended to finite temperature. We investigate the SU(2) static gluon and ghost propagators and show results for the Coulomb potential, with a focus on the Gribov ambiguity. To compute these quantities at high temperatures and to solve scaling violations we use the anisotropic Wilson gauge action.}

\FullConference{31st International Symposium on Lattice Field Theory LATTICE 2013\\
                 July 29 - August 3, 2013\\
                 Mainz, Germany}

\begin{document}

\section*{Introduction}
Zero temperature QCD in Coulomb gauge was subject of extensive studies both in the continuum formulation 
\cite{Feuchter:2004mk,Epple:2006hv} and on the lattice 
\cite{Langfeld:2004qs,Nakagawa:2007fa,Nakagawa:2008zza,Nakagawa:2008zzc,Voigt:2008rr,Burgio:2008jr,Quandt:2010yq,
Burgio:2009xp, Nakagawa:2011ar, Burgio:2012bk}. A summary of our findings was given in the talk of G.~Burgio at this 
conference \cite{Burgio:2013naa}. Recently, finite temperature was investigated in the Hamiltonian variational approach 
\cite{Heffner:2012sx} where they found a clear signal of the deconfinement phase transition in the infrared exponent of 
the ghost propagator. In the following we will extend our lattice calculations to finite temperature, in search for a 
manifestation of the deconfinement phase transition in Coulomb gauge correlation functions. 

\section{Lattice setup}
In our study we use the anisotropic SU(2) Wilson gauge action which allows us to simulate a large variety of 
temperatures (up to 6 $T_C$) at still large lattice volumes. Additionally, scaling violations in the Coulomb gauge 
propagators have been shown to get milder for large anisotropies and are expected to vanish in the lattice Hamiltonian 
limit ($\xi \rightarrow \infty$) \cite{Burgio:2012bk,Nakagawa:2011ar}.
The action is defined by 
\begin{align}
	\nonumber
	\label{eq:aniWilsonGauge}
	S[U] &= \sum_{n\in \Lambda} \left\{ \frac{\beta_s}{N_c} \sum_{i = 1}^3 \sum_{j = i+1}^{3} \real \tr \left[ 
\identity - U_{ij}(n) \right] + \frac{\beta_t}{N_c} \sum_{i = 1}^3 \real \tr \left[ \identity - U_{0i}(n) \right] 
\right\}	 \\
	\nonumber &= \frac{\beta}{N_c} \sum_{n\in \Lambda} \left\{ \frac{1}{\xi_0} \sum_{i = 1}^3 \sum_{j = i+1}^{3} 
\real \tr \left[ \identity - U_{ij}(n) \right] + \xi_0 \sum_{i = 1}^3 \real \tr \left[ \identity - U_{0i}(n) \right] 
\right\}	
\end{align}
with the inverse coupling $\beta_s = \beta/\xi_0$ in spatial directions and $\beta_t = \beta\xi_0$ in the temporal 
direction. We adjusted the bare anisotropy $\xi_0$ to correspond to a renormalized anisotropy $\xi = a_s/a_t = 4$ using 
the data from \cite{Burgio:2012bk}. For scale setting we assume a Wilson string tension of $\sigma = (440\unit{MeV})^2$.

We used lattices of size $N_t \times 32^3$ where $N_t = 128$ for $T = 0$ and $N_t \in [4,32]$ for $T > T_C$. For each 
dataset 100 configurations where generated with  heatbath and overrelaxation updates.

Coulomb gauge on the lattice is implemented by maximizing the functional
\begin{equation}
	\label{eq:gaugefunctional}
	F^g[U](t) = \frac{1}{4 N_c N_s^3} \real \sum_{x,i} \tr 
\left[g(\vec{x},t)U_i(\vec{x},t)g(\vec{x}+\hati,t)^\dagger\right]
\end{equation}
in each timeslice with respect to local gauge transformations $g(x) \in \text{SU(2)}$. Each local maximum satisfies the 
Coulomb gauge condition $\partial_i A_i = 0$. It is well known, that \eqref{eq:gaugefunctional} has very many local 
maxima which leads to the so-called Gribov problem \cite{Gribov:1977wm}. To avoid the Gribov problem we try to find the 
global maximum of \eqref{eq:gaugefunctional} by using the simulated annealing algorithm \cite{Kirkpatrick:1983zz} and 
multiple restarts of the gauge fixing procedure on random gauge transformations. For the numerical optimization we use 
a SU(2) implementation of the simulated annealing and overrelaxation algorithm on GPUs \cite{Schrock:2012fj}.

The Coulomb gauge condition is not complete: a residual space-independent gauge freedom is left unfixed. We remove this 
residual gauge freedom by the \emph{integrated Polyakov gauge} defined in \cite{Burgio:2008jr}, which is a lattice 
version of $\partial_0 \int \mass{d}{3}{\vec{x}} A_0(\vec{x},t) = 0$. Residual gauge fixing does not affect the ghost 
propagator and the Coulomb potential, since these quantities are defined at fixed time $t$. In each timeslice $t$, 
Coulomb gauge fixing is complete and any space-independent gauge transformation is just a global transformation.

\section{Gluon propagator}
In this study, we decided to use the definition of the static gluon propagator introduced in \cite{Burgio:2008jr}. 
There, a factorization of the full propagator in a function of $\absvec{p}$ and a function of $\absvec{p}/p_0$ is found 
and used to eliminate the energy dependence. On anisotropic lattices this procedure still works, but deviations from 
the factorization are found. Currently, we are investigating if this effect is only due to the larger uncertainties in 
the scale setting on the anisotropic lattices or if the simple averaging used in \cite{Nakagawa:2011ar} should be 
preferred.\footnote{The following qualitative statements about the finite temperature behaviour of the propagator is 
not affected by the definition.}

In \cite{Burgio:2008jr}, the SU(2) gluon propagator on isotropic lattices was found to satisfy the Gribov formula
\begin{equation}
	D(\absvec{p}) = \frac{1}{\sqrt{\absvec{p}^2+\frac{M^4}{\absvec{p}^2}}}.
\end{equation}
This simple ansatz does not work anymore for the propagator on anisotropic lattices. If we want to keep to a similar 
form we need to add additional parameters
\begin{equation}
      \frac{D(\absvec{p})}{\absvec{p}} = \frac{1}{\sqrt{\absvec{p}^4+\gamma M^2 \absvec{p}^2+ \alpha M^3 
\absvec{p}+M^4}}
      \label{eq:gluon:mod_gribov}
\end{equation}
where we divided by $\absvec{p}$ to get a clearer look at the IR behaviour: a non-diverging value at $\absvec{p} 
\rightarrow 0$ means a vanishing propagator.

\begin{figure}[htb]
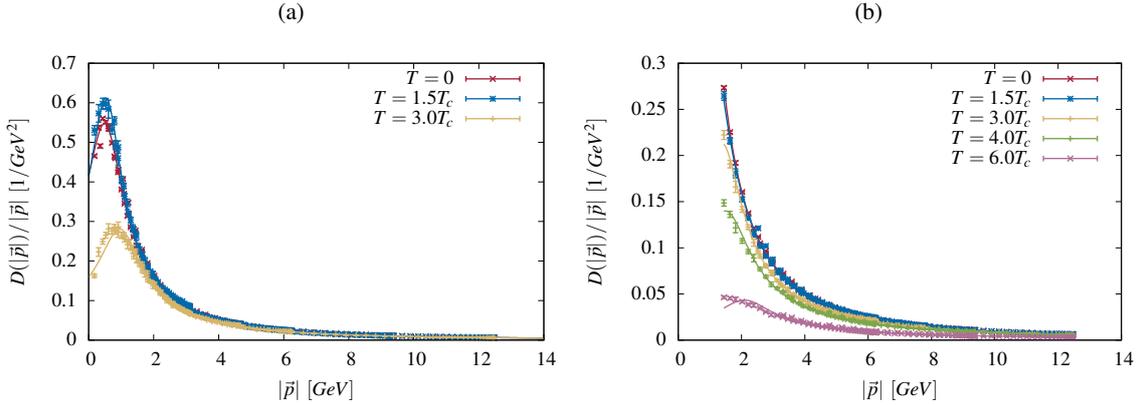

\subfloat[][]{\scalebox{.68}{\input{fig_d_upTC30_lat13proc}}}
\subfloat[][]{\scalebox{.68}{\input{fig_d_upTC60_lat13proc}}}
\caption{\label{fig:gluonprop}$D(\absvec{p})/\absvec{p}$ (a) for $T = 0,\, 1.5 T_C\text{ and } 3 T_C$ with 5 sets with 
$\beta \in [2.25,2.64]$ at each temperature.\newline(b) up to $6 T_C$ with 3 or 4 sets with $\beta \in [2.49,2.64]$}
\end{figure}

In Fig. \ref{fig:gluonprop}(a) we show $D(\absvec{p})/\absvec{p}$ at $T = 0,\, 1.5 T_C\text{ and } 3 T_C$ for $5$ sets 
of configurations with the same parameters
$\beta$ and $\xi_0$ at each temperature. To avoid spurious effects of the fitting procedure between different $T$ we 
did a combined fit to \eqref{eq:gluon:mod_gribov} with $14$ parameters: $M, \alpha \text{ and } \gamma$ for each $T$ 
and a renormalization constant for each $(\beta,\xi_0)$. The propagator at
zero temperature and at $1.5T_C$ does not show a clear difference. We expect that the small increase around the peak 
would go away at weaker coupling and with better statistics. This is also indicated by the propagator at $3 T_C$, where 
a distinct difference is now visible, but the propagator is now below the zero temperature 
propagator. This behaviour goes on towards higher temperatures (at least up to $6 T_C$), see Fig. 
\ref{fig:gluonprop}(b).

\section{Ghost dressing function}
The ghost propagator in momentum space is given by 
\begin{equation}
	G(\absvec{p}) = \frac{d(\absvec{p})}{\absvec{p}^2} = \frac{\deltafunction^{ab}}{N_s^3(N_c^2-1)} \left< 
\sum_{\vec{x},\vec{y}} \euler^{\iunit \vec{p}(\vec{x}-\vec{y})} \left[M^{-1}\right]^{ab}(\vec{x},\vec{y}) \right>
\end{equation}
where $M$ is the discrete Faddeev-Popov operator in Coulomb gauge. The ghost form factor $d(\absvec{p})$ is found to 
behave as a power $\kappa$ of the momentum $\absvec{p}$ in the infrared and has logarithmic corrections with anomalous 
dimension $\gamma$ in the ultraviolet regime:
\begin{equation}
	d(\absvec{p}) \sim \frac{1}{\absvec{p}^\kappa} \hspace{2cm} d(\absvec{p}) \sim 
\frac{1}{\log^\gamma\left(\frac{\absvec{p}}{m}\right)}.
\end{equation}
In \cite{Burgio:2012bk} an infrared exponent of $\kappa \approx 0.5$ was found by extrapolating to the Hamiltonian 
limit ($\xi \rightarrow \infty$) for the zero temperature case. This is in disagreement with findings in the 
Hamiltonian variational approach which favor $\kappa = 1$ \cite{Epple:2006hv}.
See also \cite{Burgio:2013naa}.

\begin{figure}[htb]
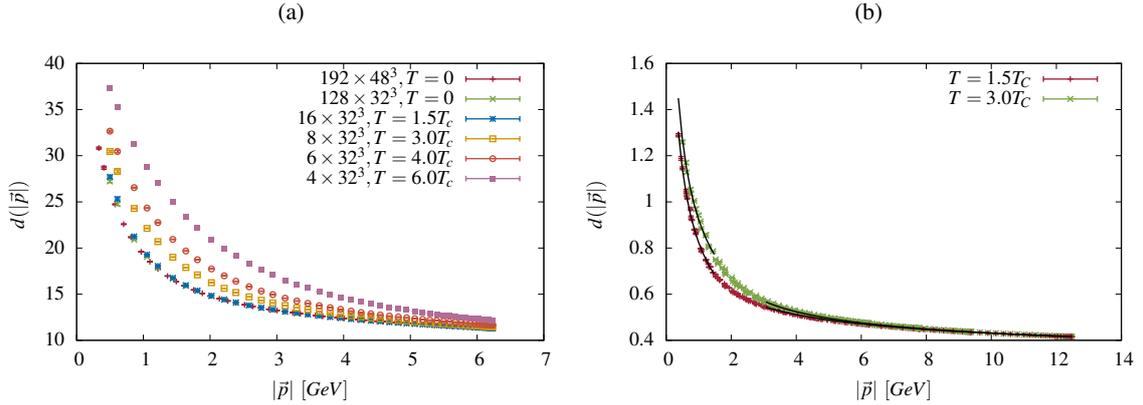

\subfloat[][]{\scalebox{.68}{\input{fig_ghost_beta24862_lat13proc}}}
\subfloat[][]{\scalebox{.68}{\input{fig_ghost_TC15_TC30}}}
\caption{\label{fig:ghost}The ghost dressing function $d(\absvec{p})$: (a) for fixed $\beta = 2.5$. (b) for $T = 1.5 
T_C \text{ and } 3.0 T_C$ from 4 or 5 sets of $\beta \in [2.40,2.64]$.}
\end{figure}
In Fig.~\ref{fig:ghost}(a) we show the ghost dressing function for fixed $\beta$. Finite volume effects are under 
control in the $128\times 32^3$ lattices as indicated by the comparison with the $192\times 48^3$ data. As in the case 
of the gluon, the ghost propagator is not sensitive to the deconfinement phase transition: we started with an IR 
exponent of about a half at $T=0$ and still have the same exponent above $T_C$. Again, starting from the $3T_C$ data 
deviations are visible. However, these deviations do not strongly affect the IR exponent which can be seen in 
Fig.~\ref{fig:ghost}(b).
At $1.5T_C$ we find an IR exponent $\kappa = 0.47$ and at $3 T_C$ we still have $\kappa = 0.46$. Both results are in 
good agreement with the zero temperature data used in \cite{Burgio:2012bk} for the extrapolation to the Hamiltonian 
limit.

In the UV, we find $\gamma = 0.63$ and $m = 0.22 \text{GeV}$ for the $1.5 T_C$ propagator. Fitting with fixed $\gamma = 
0.5$ gives the same $\chi^2/\text{d.o.f.}$ and results in $m = 0.44 \text{GeV}$. For $3 T_C$ we get $\gamma = 0.42$ 
with $m = 0.83 \text{GeV}$. Fixing $\gamma$ to 0.5 is not possible in this case, since $\chi^2/\text{d.o.f.}$ is 
doubled.

\section{Coulomb potential}
The Coulomb potential in momentum space
\begin{equation}
	V_C(\absvec{p}) = g^2\frac{\deltafunction^{ab}}{N_s^3(N_c^2-1)} \left< \sum_{\vec{x},\vec{y}} \euler^{\iunit 
\vec{p}(\vec{x}-\vec{y})} \left[M^{-1}(-\Delta)M^{-1}\right]^{ab}(\vec{x},\vec{y}) \right>
	\label{eq:coulombpot}
\end{equation}
is expected to behave like $V_C(\absvec{p}) \sim \absvec{p}^{-4}$ as $\absvec{p} \rightarrow 0$, corresponding to a 
linear rising potential in position space at large distances. However, the extrapolation to the Coulomb string tension 
$\sigma_C$
\begin{equation}
	\lim_{\absvec{p}\rightarrow 0} \absvec{p}^4 V_C(\absvec{p}) = 8 \pi \sigma_C
\end{equation}
is challenging for several reasons: (a) the conjugate gradient inversion of the operator is much more costly than the 
inversion of the ghost propagator, (b) the potential shows a large Gribov copy effect (see \cite{Voigt:2008rr}), (c) 
the extrapolation is based on very few data points in the IR (see \cite{Burgio:2012bk,Voigt:2008rr}).

With the power of our GPU gauge fixing implementation we are able to do a detailed analysis of the Gribov copy effect 
on the Coulomb potential. We fixed the gauge on 5000 random copies using simulated annealing and overrelaxation.
\begin{figure}[htb]
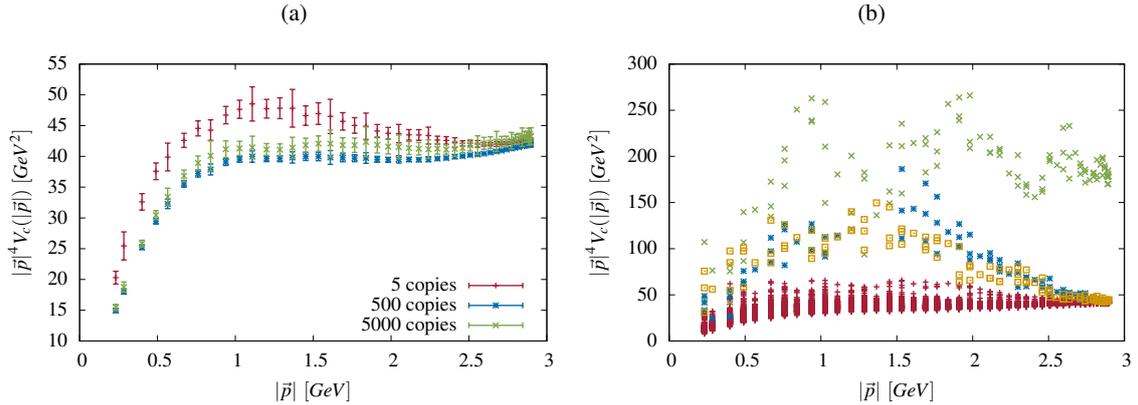

\subfloat[][]{\scalebox{.68}{\input{fig_coul_varycopies_lat13proc}}}
\subfloat[][]{\scalebox{.68}{\input{fig_coul_5000_exep_lat13proc}}}
\caption{\label{fig:coulombpot:gribov}The Coulomb potential $\absvec{p}^4 V_C(\absvec{p})$ for fixed $\beta = 2.2$ on a 
$128\times 32^3$ lattice: (a) on the best gauge copy after 5, 500 and 5000 restarts for 100 configurations. (b) the raw 
(non-averaged) data of the best copy after 5000 restarts (corresponds to the green data points on the l.h.s.).}
\end{figure}
In Fig.~\ref{fig:coulombpot:gribov}(a) we compare the potential from the best copy after 5, 500 and 5000 trials for 100 
configurations. Comparing 5, 100 (which we omitted in the plot) and 500 copies the result looks promising: the error 
bars are drastically reduced and the potential seems to converge to a stable result. However, after 5000 copies the 
error bars are again increased. The reason for this behaviour can be seen in Fig.~\ref{fig:coulombpot:gribov}(b) where 
the raw data points are plotted. The green, blue and yellow points are from three configurations which are clearly 
outliers compared to the red bulk of the remaining configurations. These outliers are from configurations for which the 
smallest eigenvalue of the operator $M(-\Delta)^{-1}M$ is more than one order of magnitude smaller compared to the 
smallest eigenvalues of the bulk. At the same time, the difference in the smallest eigenvalues of the Faddeev-Popov 
operator $M$ alone is only a factor of 2-3 and the effect on the ghost propagator from these configurations is small. 
To accommodate for these outliers in the statistics we need to incorporate much more configurations.

\begin{figure}[htb]
\centering
\scalebox{.68}{\input{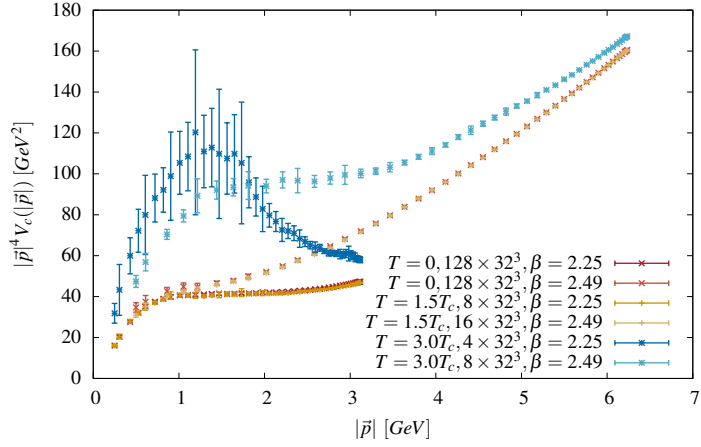}}
\caption{\label{fig:coulombpot:results}The Coulomb potential $\absvec{p}^4 V_c(\absvec{p})$ for $T = 0, 1.5T_C \text{ 
and } 3 T_C$.}
\end{figure}

Besides the Gribov copy effect, we are faced with huge scaling violations, see Fig.~\ref{fig:coulombpot:results}. In 
\cite{Voigt:2008rr} the authors already found \emph{small} scaling violations, though their study was for the gauge 
group SU(3) in a regime where discretization effects are smaller.
Again, there is no difference at $1.5 T_C$ compared to zero temperature, but at $3 T_C$ the potential changes clearly. 
The large error bars in the $3 T_C$ data is again due to one single outlier. With these problems, we are not able to 
conclude if at $3 T_C$ the Coulomb string tension changes, though very likely the string tension does not change from 
$T = 0$ to $1.5T_C$.

\section*{Conclusions and outlook}
We gave results for the gluon propagator, the ghost propagator and the Coulomb potential at finite temperature. None of 
these quantities showed a signal of deconfinement up to $1.5 T_C$. We interpret this observation as an indication that 
a naive extension of the static Coulomb gauge propagators on the lattice to finite temperature does not work. The 
propagators above $T_C$ seem to be dominated by the rising spatial string tension and insensitive to the temporal 
string tension. This is also indicated by the rise in the Gribov mass of the gluon propagator.

We also found that the computation of the Coulomb potential by the definition \eqref{eq:coulombpot} is hindered by 
several problems. Besides a strong Gribov copy effect, we find large outliers in the statistics and large scaling 
violations. With our current computational resources we cannot solve these problems satisfyingly.

We now investigate definitions of the Coulomb potential based on the temporal gluon propagator $\left< A_0 A_0\right>$ 
and partial Polyakov line correlators \cite{Marinari:1992kh,Greensite:2003xf}. Whereas the former, as a quantity 
defined at fixed timeslice, might suffer from the same problems, the latter definition seems promising for Polyakov 
lines of length $\ge 2$ lattice units.

\section*{Acknowledgments}
This work was partly supported by the Deutsche Forschungsgemeinschaft (DFG) under Contract No.~DFG-Re856/9-1 and by the 
Evanglisches Studienwerk Villigst e.V.


\end{document}